\documentclass[10pt,a4paper]{article}
\usepackage{graphicx}
\usepackage{hyperref}
\usepackage{amsmath}
\usepackage{authblk}

\title{Non-Extensive Transport Equations in Magnetized Plasmas}

\author[1]{D. S. de Oliveira}
\author[1,2]{  R. M. O. Galv\~ao}
\affil[1]{ Institute of Physics, University of S\~ao Paulo, S\~ao Paulo, Brazil}
\affil[2]{National Institute for Space Research, S\~ao Jos\'e dos Campos, S\~ao Paulo, Brazil}

\begin{document}
\maketitle

\section*{Abstract}
The transport equations for magnetized plasmas outside thermodynamic equilibrium are derived on the basis of the non-extensive statistical mechanics formulation introduced by Tsallis. The steady-state distribution function is obtained by extremization of the Tsallis entropy and the Landau collision operator is self-consistently obtained. The fluid equations and the relevant transport coefficients for electrons are then derived from the Boltzmann kinetic equation using the method of Braginskii. The results allow for consistent modelling of transport in magnetized plasmas with non-equilibrium features, in particular in the presence of long-range correlations that modify the steady-state distribution function, on the basis of only one parameter, q. Since many transport coefficients can be measured in a single experiment and they all depend on q, the determination of its proper value to fit specific experimental results, using the present model, can be quite robust. We also apply the model introduced here in the transport of the heat in the solar wind and in the interpretation of the cold pulse in magnetic confinement plasmas.

\section{Introduction} \label{sec:intro}

A common distinctive feature of most laboratory and space plasmas is that of being far from thermodynamic equilibrium. In the turbulent state usually observed, in particular when the underlining plasma modes have long-range correlations and/or relevant wave-particles interactions, the particle velocity distribution functions are quite far from a Maxwellian, with long energetic tails, especially for electrons, but also for another species in some cases \cite{Demars90,Izacard2016}.

One of the first observations of long-tail electron distribution functions was made by Vasyliunas, when analyzing low-energy electron fluxes in the magnetosphere measured by the OGO1 and OGO3 satellites \cite{Vasyliunas1968}. In order to explain the data, an empirical model distribution function approaching a power law at high energies was introduced, which became known as the $\kappa$-distribution function. Recent data from STEREO Solar Terrestrial Probes Program show clearly long tails in the quite time superhalo electron velocity distribution function, which are well modelled by the $\kappa$-distribution function \cite{Wang2012}. Due to its wide range of applications, $\kappa$-distribution functions have been frequentçly applied investigated in many plasma studies over the years, in particular regarding the modification of the dispersion relation of different kinetic plasma modes and the evolution  of nonlinear instabilities \cite{Summers1991,Mace1995}. This empirical distribution functions was explained in terms of the Tsallis statistics, a generalization of the Boltzmann statistics for systems with long-range correlation between particles \cite{Leubner2002}.

In tokamak plasmas, long-tail electron distributions have been observed over the whole plasma column (edge, confinement region, and core) and related to several different mechanisms, such as magnetic reconnection, high-energy ions, non-local electron transport, neutral ionization, and plasma heating \cite{DelaLuna2003,Jaworski2013,Popov2015,DuBois2018}. However, in most of the cases, without a proper explanation or even a self-consistent model. Investigations of the plasma dynamics suggested that high energetic electrons, known as suprathermal electrons, seem connected with several instabilities in magnetic confinement plasmas, for instance, internal kink modes, sawtooth instabilities, electron fishbones \cite{Wong2000,TcvTeam2015}. Therefore, the understanding and theoretical formulation, from predictive models, of the effect of such particles on plasma transport are important in order to improve the operational conditions and design of the magnetic plasma devices. 

The major theoretical issue regarding the descriptions of non-Maxwellian distributions is the absence of a first-principles model. Indeed, a self-consistent model without the usual limitations of the Boltzmann-based fluid theory is not available \cite{Izacard2017}. In order to overcome this difficulty, alternative fluid equations based on different types of long-tail distribution functions have been lately considered to model various plasma systems, especially in numerical simulations \cite{Izacard2016}. In particular, one scheme is to approximate the distribution function by a series of Maxwellian distributions, with the coefficients of proportionality determined by numerically fitting the experimental data \cite{Izacard2016}. It can be shown that this approach can be asymptotically approximated by the model described in this work. However, despite their success in recovering some quantitative experimental results, the basic reasons for the validity of the long-range distributions remains somewhat an open issue in the plasma literature. 

An entirely different theoretical framework to model physical systems outside thermodynamic equilibrium with long-range correlations was pioneered by Tsallis \cite{Tsallis1988,Tsallis2009}. In his approach, a generalized form of the Boltzmann entropy is introduced, depending on a single free parameter, dubbed $q$. Although this parameter can, in principle, be formally obtained from basic physics, practically such task reveals practically impossible and, in general, $q$ is introduced as a fitting parameter. Many applications of the Tsallis theory have been reported in plasma physics; for instance, the equilibrium density after turbulent relaxation in a pure electron plasma \cite{Antenodo1997}, superdiffusion transport in dusty plasmas \cite{Liu2008}, plasma oscillations \cite{Lima2000}, $\kappa$-distribution functions as first principle distributions in Tsallis theory ($\kappa=1/(q-1)$) \cite{Livadiotis2009}, transport coefficients in the BGK collisional approximation \cite{Boghosian1999,Bezerra2003}, vorticity distribution at the plasma edge of tokamaks \cite{Golcalves2018}, etc. The main advantage of Tsallis method is that it self-consistently generalizes both the distribution function and the underlying statistics, namely, non-extensive statistics (or q-statistics), such that many physical relevant quantities can be derived starting from entropy only \cite{Tsallis2009}.

The foundations of the q-statistics are based upon the generalized entropy \cite{Tsallis1988,Tsallis2009},
\begin{equation}
S_q = \frac{1-\sum_{\mu} p_\mu^q}{q-1},
\label{qentropy}
\end{equation}
where we took $k_B=1$, which is equivalent to measure the temperature in energy units; $q$ is a real number, where for $q\to1$, the above expression recover the Boltzmann entropy ($S_B$); $p_\mu$ is the normalized probability of the state $\mu$; and the generalized mean (or q-mean) value, which preserves the statistical properties of average value, is
\begin{equation}
O_q = \frac{\sum_\mu p_\mu^q O_\mu}{\sum_\mu p_\mu^q},
\label{qmean}
\end{equation}
where $O_\mu$ is the operator associated to the mean value $O_q$.

From a theoretical point of view, the adoption of the non-extensive entropy approach to model non-Maxwellian distributions functions is justified by it being one of most robust generalizations of the Boltzmann entropy available in the literature. Indeed, it shares all desirable properties for a ``good'' entropic definition with Boltzmann entropy, $S_B$ \cite{Tsallis2009}. The main difference between $S_q$ and $S_B$ is the non-additivity property if $q\neq1$, which lead, for some time, to the misconception of the Tsallis entropy being a ``non-extensive entropy''. However, additivity is a sufficient condition rather than a necessary one. The extensiveness of any entropy is a consequence of the factorization of the likelihood function of independent states (or systems), which may or may not result on the extensiveness of $S_q$, in the same way as for $S_B$ \cite{Tsallis2015}. Nevertheless, we will keep the ``non-extensive entropy'' terminology that became widespread in the literature.

In this paper, starting only from definition of $S_q$, the closed electron fluid equations in the limit of weak interactions will be derived from a self-consistent non-extensive kinetic theory (q-kinetic theory). We restrict the analysis to the electron fluid equations in plasmas with only one ionic component, for the sake of simplicity. Indeed, this allows to use the electron-ion mass ratio as an expansion parameter and it makes unnecessary the evaluation of the stress tensor. Therefore, this work has to be considered as a first step in the development of a plasma transport model based on the Tsallis entropy, which will be extended to include the ion fluid equations. In section II, the continuous formulation of the q-kinetic theory as well as the temperature definition are discussed. The collisional operator is found in Section III with the help of the Kinetic Interaction Principle (KIP) \cite{Kaniadakis2002}, whereas the general aspects of the q-kinetic model and the Chapman-Enskog method \cite{chapman,Braginskii1958,cercignani1990} are presented in Section IV and V, respectively. Section VI is dedicated to the numerical evaluation of the main transport coefficients. Two applications in space and tokamak plasmas are presented in Section VII.

The foundations of the q-statistics are based upon the generalized entropy \cite		{Tsallis1988,Tsallis2009},
\begin{equation}
S_q = \frac{1-\sum_{\mu} p_\mu^q}{q-1},
\label{qentropy}
\end{equation}
where we took $k_B=1$, which is equivalent to measure the temperature in energy units; $q$ is a real number; $p_\mu$ is the normalized probability of the state $\mu$; and the generalized mean (or q-mean) value, which preserves the statistical properties, is
\begin{equation}
O_q = \frac{\sum_\mu p_\mu^q O_\mu}{\sum_\mu p_\mu^q},
\label{qmean}
\end{equation}
where $O_\mu$ is the operator associated to the mean value $O_q$.

From a theoretical point of view, the adoption of the non-extensive entropy approach to model non-Maxwellians distributions functions is justified by it being one of most robust generalizations of the Boltzmann entropy ($S_q$) available in the literature. Indeed, it shares all desirable properties for a ``good'' entropic definition with Boltzmann entropy, $S_B$ \cite{Tsallis2009}. The main difference between $S_q$ and $S_B$ is the non-additivity property if $q\neq1$, which lead, for some time, to the misconception of the Tsallis entropy being a ``non-extensive entropy''. However, additivity is a sufficient condition rather than a necessary one. The extensiveness of any entropy is a consequence of the factorization of the likelihood function of independent states (or systems), which may or may not result on the extensiveness of $S_q$, in the same way as for $S_B$ \cite{Tsallis2015}. Nevertheless, we will keep the ``non-extensive entropy'' terminology that became widespread in the literature.

In this paper, starting only from definition of $S_q$, the closed electron fluid equations in the limit of weak interactions will be derived from a self-consistent non-extensive kinetic theory (q-kinetic theory). We restrict the analysis to the electron fluid equations in an one-component plasma for the sake of simplicity. Indeed, this allows to use the electron-ion mass ratio as an expansion parameter and it makes unnecessary the evaluation of the stress tensor. Therefore, this work has to be considered as a first step in the development of a plasma transport model based on the Tsallis entropy, which will be  extended to include the ion fluid equations. In section II, the continuous formulation of the q-kinetic theory as well as the temperature definition are discussed. The collisional operator is found in Section III with the help of the Kinetic Interaction Principle (KIP) \cite{Kaniadakis2002}, whereas the general aspects of the q-kinetic model and the Chapman-Enskog method \cite{chapman,Braginskii1958,cercignani1990} are presented in Section IV and V, respectively. Section VI is dedicated to the numerical evaluation of the main transport coefficients. Two applications in space and tokamak plasmas are presented in Section VII.

\section{Continuous formulation and temperature definition}\label{sec:cont}

In the continuous formulation, the generalized entropy is defined by \cite{Tsallis2009}
\begin{equation}
S_q = \int d\mathbf{v}\,\frac{p -p^q}{q-1}, \label{contsq}
\end{equation}
where $k_B=1$ (temperature measured in energy units), $q$ is a real number, $p$ is the normalized distribution function, the integrals extend all over the velocity phase space, and the q-mean is defined by
\begin{equation}
O_q = \frac{\int d\mathbf{v} \, O(\mathbf{v}) p^q}{\int d\mathbf{v} p^q}. \label{contqmean}
\end{equation}

In analogy, the normalization condition and internal energy of the plasma particles become
\begin{eqnarray}
n = \int d\mathbf{v} p, \label{contnorm}
\\\nonumber
\\
\frac{u_q}{n} = \frac{\int d\mathbf{v} \left( \frac{m v^2}{2} + e_a \phi\right) p^q}{\int d\mathbf{v}\,p^q}, \label{contintener}
\end{eqnarray}
where $e_a$ is the electric charge of the particle species, $m$ is the mass, $\phi$ is the electric potential and the index ``a'', which distinguishes electrons and ions, has been suppressed in quantities but the charge, since the calculations in this section are identical for all species.

Here, it is important to notice that Eq.(\ref{contsq}) is defined up to a constant in the power law of $p$. However, when obtaining the distribution function by the standard variational extremization procedure, as we present next, such constant can be conveniently coupled in the Lagrange multipliers and, therefore, disappears from final expression.

From the standard variational extremization procedure of the Lagrangian of the entropy \cite{Tsallis1998,Martinez2000,Bagci2009}, with the constrains given by Eq.(\ref{contnorm}) and Eq.(\ref{contintener}), the equilibrium distribution function is obtained
\begin{equation}
p_0 = \beta_n \left[ 1- (1-q) \beta_u \left( \frac{ mv^2}{2} + e_a \phi - \frac{u_q}{n}\right)\right]^{\frac{1}{1-q}}, \label{ordfunequi}
\end{equation}
where $\beta_n$ and $\beta_u$ account for the Lagrange multipliers of the normalization constants (density) and internal energy (temperature dependent), respectively. The above distribution function presents itself as the  well know power-law equilibrium distribution function of the q-statistics, which replaces the ordinary Maxwellian distribution in the traditional approach. Henceforward, we limit our analysis to $q>1$, where long-tails distributions are found (actually, for $q<1$, $p$ has an upper limit in velocity space given by the Tsallis cut-off, which limits the distribution function \cite{Tsallis2009}).

It is also convenient for our purposes the formulation in terms of escort distribution functions $f= n p^q/\left( \int d\mathbf{v}\, p^q\right)$ (q-escort distribution) \cite{Tsallis2009,beck1995}, which recover the ordinary statistical average in q-mean, as can be verified in Eq.(\ref{contqmean}). In this new formulation, Eq.(\ref{ordfunequi}) is rewritten as
\begin{equation}
f_0 = n_0 \left(\frac{ m \beta_q}{2}\right)^{\frac{3}{2}} \left[ 1-(1-q) e_a \beta_q \phi\right] A_q \left[ 1-(1-q) \beta_q \left( \frac{m v^2}{2} + e_a \phi\right)\right]^{\frac{q}{1-q}}, \label{distequi}
\end{equation}
where the normalization constant $A_q$, obtained from Eq.(\ref{contnorm}), is
\begin{equation}
A_q= \left\{
\begin{array}{ll}
\pi^{-\frac{3}{2}}, & q=1;\\
\frac{(q-1)^{\frac{1}{2}}}{\pi^{\frac{3}{2}}} \frac{\Gamma\left(\frac{1}{q-1}\right)}{\Gamma\left( -\frac{1}{2} + \frac{1}{q-1}\right)}, & 1<q<3,
\end{array}\right.\label{normcte}
\end{equation}
and we also have used $n(\phi)$ and $u_q$ obtained from Eqs.(\ref{contintener}) and (\ref{contnorm}) with the substitution of Eq.(\ref{distequi}),
\begin{eqnarray}
\frac{n}{n\left(\phi=0\right)} &=& \frac{n}{n_0} = \left[1-(1-q) e_a \beta_q \phi\right]^{\frac{3}{2}+\frac{1}{1-q}}, \quad 1-(1-q)e_a \beta_q \phi >0; \label{meann}
\\
u_q &=& \frac{2}{5 -3q} \frac{n}{\beta_q} \left( \frac{3}{2} +e_a \beta_q \phi\right); \quad 1<q< \frac{5}{3},\label{meanintener}
\end{eqnarray}
where the upper limits on $q$ correspond to the maximum values for which the integrals of $f_0$ diverges, and $\beta_q$ is given by
\begin{equation}
\beta_q = \frac{ \beta_u n\left[ \int d\mathbf{v}\, p^q\right]^{-1}}{1 + (1-q) u_q\beta_u \left[\int d\mathbf{v}\,p^q\right]^{-1}}. \label{betaq}
\end{equation}

The connection with thermodynamics is set by the temperature definition, which is not unique in q-statistics as it is in the Maxwell-Boltzmann (MB) theory. In fact, the kinetic and the equilibrium temperatures are essentially different from the Lagrangian temperature; they even may have different physical interpretations \cite{Nobre2015,Abe1999,Livadiotis2014,Ruseckas2016}. In the presented model, the temperature is defined by the generalized zeroth law \cite{Ruseckas2016,Abe2001}
\begin{equation}
\left( \frac{\partial S_q}{\partial u_q} \right)_n \left[ 1+(1-q) S_q/n\right]^{-1} = \frac{1}{T}, \label{gen0law}
\end{equation}
where $T$ is the equilibrium temperature of the system measured by a thermometer. This is convenient for our purpose because, by substituing Eq.(\ref{contsq}), Eq.(\ref{betaq}) and Eq.(\ref{gen0law}) into Eq.(\ref{meanintener}), we recover the classical internal energy $u_q = \frac{3}{2}n T + n e_a\phi$, which allows us to understand  $T$ as the usual average kinetic energy as well as the terms of the energy balance equation to be derived in Sec.IV. From Eq.(\ref{contsq}), (\ref{betaq}), and (\ref{gen0law}), we also define the auxiliary temperature
\begin{equation}
\frac{1}{\beta_q} = T_q = \frac{ 5 -3q}{2} T + (1-q) e_a \phi. \label{tq}
\end{equation}

If the density given in Eq.(\ref{meann}) is expanded at $q\to1$ and the weak interactions condition (namely, $e_a \phi/T\ll1$) is applied, i.e, only collisional transport, then
\begin{equation}
n = n_0 e^{-\frac{e_a \phi}{T}} \left( 1-\frac{1-q}{2} \left(\frac{e_a \phi}{T}\right)^2 +\ldots\right)\approx n_0 e^{-\frac{e_a \phi}{T}}, \label{nboltzmann}
\end{equation}
recovering the ordinary expression of the density from Boltzmann statistics. In turn, the Debye length and, therefore, the upper cut-off of the collision cross section do not change \cite{Hazeltine1998,Helander2005}. An analogous expansion of the distribution function in Eq.(\ref{distequi}) around $q\to1$ yields a series of coefficients multiplied by Maxwellians, suggesting that the numerical expansions aforementioned may asymptotically approach q-distributions.

The weak interaction condition is rigorously verified for $q\in [1,1.4]$, where $2 (q-1)/(5 -3 q)\leq 1$ guarantees that the second term in Eq.(\ref{tq}) is always smaller than the first. This restriction is needed because when $q\to5/3$, the dependence on $T$ in Eq.(\ref{distequi}) is negligible and, therefore, the width of the distribution is set only by $\phi$. In this circumstance, any fluctuation of $\phi$, however small compared to $T$, is noticeable; even if $e_a \phi/T\ll1$. This feature of the q-distributions is quite interesting for turbulent statistical descriptions, in which the transport due potential fluctuations are more important than that due collisions \cite{balescu2005}. Finally, applying $e_a \phi/T\ll1$ and using the self-referential property of the escort distributions \cite{Tsallis2009}, the expressions for the distribution function (Eq.(\ref{distequi})) and $T_q$ (Eq.(\ref{tq})) yield
\begin{equation}
f_0\approx n \left(\frac{m}{2 T_q}\right)^\frac{3}{2} A_q \left[1-(1-q) \frac{ m v^2}{2 T_q}\right]^\frac{q}{1-q};\quad T_q \approx \frac{ 5 -3 q}{2} T. \label{distNtq}
\end{equation}

These approximations seem to be compatible with the numerical simulations of the observed superhalo electron velocity distribution function \cite{Yoon2012} and the electron temperature measurements in solar winds \cite{Nicolaou2016}.

\section{q-Landau Operator}\label{sec:qlandauop}

The classical Boltzmann equation is derived assuming only the hypotheses of statistical independence between collisions, the known molecular chaos \cite{chapman,cercignani1990}. Evidently, in the context of the q-statistics, the complete statistical uncorrelation between collisions for all states at all times is not applicable, because of the long-range correlations, a basic hypothesis of the Tsallis entropy \cite{Tsallis1988}. Furthermore, the current generalizations of the molecular chaos for q-statistics leading to extensions of the Boltzmann equation \cite{Lima2001,Chavanis2004,Abe2009} cannot be consistently employed for obtaining neither the fluid equations from the kinetic equation nor the correct expression for the collisional operator in non-thermal plasmas without further hypotheses. This problem is addressed in this section. For this, the Kinetic Interaction Principle (KIP) method, introduced in Ref.cite{Kaniadakis2001}, is employed to derive the q-kinetic equation. This has been motivated by Ref.\cite{Chavanis2004}, where the first generalization of the Landau operator was presented for normal q-distributions (Eq.(\ref{ordfunequi})). Here, the generalized kinetic equation is derived for q-escort distributions, since in this formulation the determination of the fluid equations from the kinetic theory follows the standard kinetic moments procedure, because of the already mentioned recovery of the standard statistical average.

The KIP method states that the collisional evolutions of the distribution function in phase space is \cite{Kaniadakis2002,Kaniadakis2001}
\begin{equation}
\frac{df}{dt} = \int d\mathbf{v}'d\mathbf{v}_1 d\mathbf{v}_1'\,\left[ \Pi\left( \mathbf{r},\mathbf{v}'\to\mathbf{v},\mathbf{v}_1'\to\mathbf{v}_1,t\right)-\Pi\left(\mathbf{r},\mathbf{v}\to\mathbf{v}',\mathbf{v}_1\to \mathbf{v}_1',t\right)\right], \label{kipeq}
\end{equation}
where $\mathbf{r}$ is the position where collision occurs, $d/dt$ is the convective derivative, $\mathbf{v}$, $\mathbf{v}'$, $\mathbf{v}_1$ and $\mathbf{v}_1'$ are, respectively, the incident and target velocity of the particles before and after the collision, and $\Pi$ is the probability of transitions. $\Pi$ is further decomposed as a generic combination of positive definite functions
\begin{equation}
\Pi = \mathrm{Tr} \left(\mathbf{r},\mathbf{v}',\mathbf{v},\mathbf{v}_1',\mathbf{v}_1,t \right) \gamma\left(f,f'\right) \gamma\left(f',f\right), \label{probtran}
\end{equation}
where $\mathrm{Tr}$ is the transition rate and $\gamma(f,f')=a(f) b(f')c(f,f')$, with $a$, $b$ and $c$ being positive functions, in which $c(f',f) =c(f,f')$ accounts for the influence of the populations on the collision process. In explicit terms of the $a$, $b$ and $c$ functions, Eq.(\ref{kipeq}) is
\begin{equation}
\frac{df}{dt} = \int d\mathbf{v}' d\mathbf{v}_1 d\mathbf{v}_1 \, \mathrm{Tr}(r,\mathbf{v}',\mathbf{v},\mathbf{v}_1',\mathbf{v}_1,t) c c_1 \left[ a' b a_1' b_1 - a b' a_1 b_1'\right],
\end{equation}
where $a=a(f)$ and $a'=a(f')$ and so on.

In the weak interaction condition, the collisions are binary and cause only small changes on the particle velocities $|\mathbf{\Delta}| = |\mathbf{v} - \mathbf{v}_1|\ll (|\mathbf{v}|,|\mathbf{v}_1|)$, and functions $a$ and $b$ in Eq.(\ref{probtran}) can be expanded as power series. The general steps of the calculations can be found in Ref.\cite{Chavanis2004}; the resulting kinetic equation is
\begin{equation}
\frac{df}{dt} = \frac{\partial }{\partial v_\mu} \int d\mathbf{v}_1 K_{\mu\nu} \frac{c c_1}{m} \left[ \frac{ g_1 h}{m} \frac{\partial f}{\partial v_\mu} - \frac{ g h_1}{m_1} \frac{\partial f_1}{ \partial v_{1\mu}}\right],
\label{ghop}
\end{equation}
where
\begin{equation}
g = ab; \quad h = \frac{d a}{d f} b - a \frac{d b}{d f}; \quad K_{\mu\nu} = \int d\mathbf{v}_1 \mathrm{Tr}( \mathbf{r}, \mathbf{v}, \mathbf{v}_1,t; \mathbf{\Delta}). \label{hgNK}
\end{equation}

The relations for $g_1$ and $h_1$ are analogous and the indices $\mu$ and $\nu$ stand for the Cartesian coordinates $x$, $y$ and $z$. The tensor $K_{\mu\nu}$ depends on the cross section of the collisions and it is calculated from the standard Newton mechanics \cite{Hazeltine1998} as
\begin{equation}
K_{\mu\nu} = \frac{ 2 \pi e^2 e_1^2  \lambda}{ m } U_{\mu\nu} = \frac{ 2 \pi e^2 e_1^2  \lambda}{m} \frac{ \delta_{\mu\nu} u^2 - u_\mu u_\nu}{ u^3}, \label{expK}
\end{equation}
where $m$ is the incoming particles mass, $u_\mu = v_\mu-v_{1\mu}$ is the relative velocity, $e$ and $e_1$ are the charges of the particles involved in the binary collision, and $\lambda = \ln(\lambda_D/r_{imp})$ is the Coulomb logarithm with $\lambda_D$ the Debye length and $r_{imp}$ the impact parameter.

The functions $h$ are obtained from $G'' = h/g$, where $S_q = \int d \mathbf{v} G(f) $ \cite{Kaniadakis2002,Kaniadakis2001}. In order to define all functions uniquely, the $c$ and $g$ functions have to be chosen properly. Since the collisions are binary and the Coulomb force is symmetric, the instantaneous process is independent of the particle populations and, therefore, $c c_1 =1$. In the weak interactions limit, the integrals in Eq.(\ref{ghop}) must approach a diffusive process in phase space \cite{landau2008}. Accordingly, the choice $g=f$ ($g_1=f_1$) enables interpreting these integrals as the $h$ ($h_1$) weighted average of the momentum transfer from $f$ to $f_1$ (or $f_1$ to $f$); we emphasize that this interpretation is only possible because of the q-mean in Eq.(\ref{qmean}) being changed to the ordinary form in the q-escort approach. The same expressions were found in Ref.\cite{Chavanis2004}; however, as simple mathematical choices justified only by identification of the classical Landau collision operator for $q\to1$.

The final expression of the collisional operator in our model is
\begin{equation}
C(f,f_1) = \frac{ 2 \pi e^2 e_1^2  \lambda}{m} \frac{\partial }{\partial v_\nu} \int d\mathbf{v}_1 U_{\mu\nu} \left[ \frac{f_1}{m} \frac{\partial f^*}{\partial v_\mu} - \frac{f}{m_1} \frac{\partial f_1^*}{\partial v_{1\mu}} \right], \label{qlandop}
\end{equation}
where $f^*$ (or $f_1^*$) is
\begin{equation}
f^* = \frac{ n f^\frac{1}{q}}{q k_q} \approx \frac{ 5 -3q}{2} n \left( \frac{m}{2 T}\right)^\frac{3}{2} A_q \left[ \frac{f}{n \left(\frac{m}{2 T_q}\right)^\frac{3}{2} A_q}\right]^\frac{1}{q}, \label{fstar}
\end{equation}
and  we have used, in advance, that the solutions of interest are $f=f_0+\delta f$, where $\delta f$ is the first order solution and $\delta f/f_0 \ll 1$, which allows $k_q(f)\approx k_q(f_0) = \int d\mathbf{v} f_0^{1/q}/n$.

In Eq.(\ref{qlandop}), the parameter $q$ does not appear explicitly and no further hypothesis besides the particles being charged were made. Therefore, the extension of the results so far obtained to all particle species in the plasma, whose populations are described by different distribution functions, with different parameters or $q$'s, is straightforward,
\begin{equation}
C_a =\sum_b C_{ab} = \sum_b \frac{ 2 \pi e_a^2 e_b^2 \lambda_{ab}}{m_a} \frac{\partial}{\partial v_{a\nu}} \int d\mathbf{v}_b U_{\mu\nu} \left[ \frac{f_b}{m_a} \frac{\partial f_a^*}{\partial v_{a\mu}} -\frac{f_a}{m_b}\frac{\partial f_b^*}{\partial v_{b\mu}}\right];
\label{qlandopab}
\end{equation}
and the non-extensive multicomponent plasma kinetic equation in terms of q-escort distributions is
\begin{equation}
\frac{\partial f_a}{\partial t} +\mathbf{v}_a \cdot \nabla f_a + \frac{\mathbf{F}_a}{m_a} \cdot \nabla_{\mathbf{v}_a} f_a = C_a,
\label{finalkineq}
\end{equation}
where $\mathbf{v}_a$ is the velocity of the ``$a$'' species and $\mathbf{F}_a$ is the external force acting on these.

The proof of the constraint relations (conservation of mass, momentum, and energy) as well as the H-theorem is given in Ref.\cite{Kaniadakis2002,Chavanis2004,Kaniadakis2001} in general terms, i.e., before the supposition of a specific statistics, including Tsallis statistics. Further properties of the q-Landau operator are shown in Appendix A.

\section{Kinetic Model}\label{sec:kinmodel}
\subsection{Electron-Ion Approximations}

Due the mass disparity between electrons and ions, $m/m_i\ll 1$, the velocity of the ions is, in general, much smaller than that of the electrons. This condition enables the expansion of $U_{\mu\nu}$ in power series of the ion velocity
\begin{equation}
U_{\mu\nu} = V_{\mu\nu} -\frac{\partial V_{\mu\nu}}{\partial v_{\xi}} v_{i\xi} + \frac{1}{2} \frac{\partial^2 V_{\mu\nu}}{\partial v_\zeta \partial v_\xi} v_{i\xi} v_{i\eta}, \label{exptensor}
\end{equation}
where $V_{\mu\nu} = U_{\mu\nu}(\mathbf{v}_i=0)$ and we took, for convenience, the coordinate system where $\mathbf{V}_i =0$. Then, the above expression can substituted in Eq.(\ref{qlandopab}) and integrated over $\mathbf{v}_i$, with the boundary condition ${f(v\to\infty)=0}$, yielding the approximative expression of the electron-ion collision operator 
\begin{equation}
C_{ei} =\frac{ 2 \pi e^2 e_i^2 n_i \lambda}{m} \frac{\partial}{\partial v_\nu} \left[ V_{\mu\nu} \frac{ \partial f^*}{\partial v_\mu} - \frac{m}{m_i} \left( 2 \frac{v_\mu}{v^3} \frac{n_i^*}{n_i} f + \frac{ 3 v_\mu v_\nu -v^2 \delta_{\mu\nu}}{v^5} \frac{T_i}{m} \frac{\partial f^*}{\partial v_\mu}\right)\right]
\label{opei}
\end{equation}
where $n_i^*=\int d\mathbf{v}_i\, f_i^*$. Neglecting terms of $\mathcal{O}\left(m/m_i\right)$, the principal part of the $C_{ei}$ is
\begin{equation}
C_{ei}'(f)= \frac{2 \pi e^2 e_i^2 n \lambda}{m} \frac{\partial}{\partial v_\nu} \left[ V_{\mu\nu} \frac{\partial f^*}{\partial v_\mu}\right]
\label{prinopei}
\end{equation}
where the local neutrality $n\approx n_i$ was invoked and, except by $f^*$, the above expression is equal to the classical operator \cite{Braginskii1958,Braginskii1965}. Furthermore, it can be also verified that the same expression for the e-i collision frequency for the zeroth collision classical operator is held \cite{Helander2005}, namely,
\begin{equation}
\omega_{ei} = \frac{3 \pi^\frac{1}{2}}{4 \tau}\left(\frac{v_T}{v}\right)^3, \label{freqcolei}
\end{equation}
where $v_T^2 = 2 T/m$ is the thermal velocity and $\tau$ is the relaxation time defined as
\begin{equation}
\tau = \frac{ 3 \sqrt{2} T^\frac{3}{2}}{4 \sqrt{ 2 \pi} e^2 e_i^2 n  \lambda} \label{taurelax}.
\end{equation}

\subsection{Transport Equations}

From Eq.(\ref{finalkineq}), it follows that, for a fully ionized single species plasma, in the presence of stationary electromagnetic fields, the kinetic equation in terms of the peculiar velocity of the electrons, $\mathbf{v}=\mathbf{v}'-\mathbf{V}$ (the velocity of the electrons is now $\mathbf{v}'$), is
\begin{equation}
\frac{ d f}{d t} + \mathbf{v}\cdot \nabla f + \left( \frac{e}{m} \left( \mathbf{E}' +\mathbf{v}\times \mathbf{B}\right) - \frac{d \mathbf{V}}{d t}\right) \cdot \nabla_v f = C_e,
\label{eqkinpecv}
\end{equation}
where $f$ is the electron distribution function, $\mathbf{E}'=\mathbf{E}+\mathbf{V}\times\mathbf{B}$, $\mathbf{E}$ and $\mathbf{B}$ are, respectively, the electric and magnetic fields in the laboratory frame, and $C_e =C_{ee}(f)+C_{ei}(f)$ is the total collisional operator accounting for electron-electron collisions (e-e collisions) and electron-ion collisions (e-i collisions). Since the e-i collision operator in Eq.(\ref{prinopei}) is independent of the distribution function of the ions, the above equation does not depend explicit on $f_i$. Therefore, the evolution of the distribution function of the electrons can be obtained independently of the evolution of $f_i$ as well as its fluid equations.

In the weak interaction limit, only the first three fluid moments of the kinetic equation are enough for a reasonable approximation of the fluid equations \cite{cercignani1990}. Since the q-escort approach holds the ordinary statistical average and Eq.(\ref{eqkinpecv}) has the exact form of the kinetic equation for the Maxwell-Boltzmann statistics \cite{Helander2005,Braginskii1965}, the first three moments (namely, multiplying the kinetic equation by each of $(1,m\mathbf{v}, mv^2/2)$ and integrating over $\mathbf{v}$) recover the classical transport equations system
\begin{eqnarray}
\frac{d n}{d t} +n \nabla\cdot \mathbf{V} = 0; 
\label{conteq}
\\
n m\frac{ d \mathbf{V}}{d t} + \nabla p = e n\left(\mathbf{E}' + \mathbf{V}\times \mathbf{B}\right) + \mathbf{R};
\label{moveq}
\\
\frac{3}{2} n \frac{ d T}{d t} + p \nabla \cdot \mathbf{q} = Q,
\label{baleq}
\end{eqnarray}
where the following quantities have been introduced
\begin{equation}
\begin{array}{ccc}
\displaystyle p = \int d\mathbf{v}' \frac{ m {v'}^2}{2} f = n T;&\quad&\displaystyle Q= \int d\mathbf{v}' \frac{ m {v'}^2}{2} C_e;
\\
\displaystyle\mathbf{q} = \int d\mathbf{v}' \frac{m {v'}^2}{2} \mathbf{v}' f; &\quad& \displaystyle \mathbf{R} = \int d\mathbf{v}'\, m\mathbf{v}' C_{ei}
\label{defflux}
\end{array}
\end{equation}
the hydrostatic pressure $p$, the heat flux $\mathbf{q}$, the friction force $\mathbf{R}$, and the thermal energy transfer $Q$. As already mentioned, since we are considering only electrons, the viscosity tensor is neglected in these equations. The Eqs.(\ref{conteq}-\ref{defflux}) are a closed system of the fluid equations when $\mathbf{q}$ and $\mathbf{R}$ are given in terms of the plasma parameters, which requires the explicit solution of Eq.(\ref{eqkinpecv}).

\subsection{Zero Order Friction Force}

If the disturbance caused by the ions on the electron velocity is small, the displacement on the electron equilibrium distribution function is of the  order of $\mathbf{U}= \mathbf{V}-\mathbf{V}_i$ \cite{Braginskii1965}. For such shift, the relative velocity $\mathbf{u}$ of the e-i collision operator is independent of $\mathbf{V}_i$ and, therefore, the zeroth order friction force $\mathbf{R}^{(0)}$ can be calculated from Eq.(\ref{prinopei}). From Eq.(\ref{defflux}), knowing that the small perturbation corresponding to a small shift in comparison with the thermal electron energy ($U/\sqrt{T/m} \ll 1$), which, in turn, allows the expansion of $f_0$ in power series of $\mathbf{U}$, the friction force become

\begin{equation}
\mathbf{R}^{(0)} = \frac{ 2 \pi e^2 e_i^2 n  \lambda}{m} \int d\mathbf{v} \,m \mathbf{v} \frac{\partial}{\partial v_\nu} \left[ V_{\mu\nu} \frac{\partial f_0^*(\mathbf{v} - \mathbf{U})}{\partial v_\mu} \right] = - \sqrt{\frac{2}{5 -3 q}} \frac{A_q \pi^\frac{3}{2}}{q} \frac{m n}{\tau} \mathbf{U}, \label{fric0}
\end{equation}
where $\tau$ is the relaxation time given in Eq.(\ref{taurelax}).

\begin{figure}
  \centerline{\includegraphics{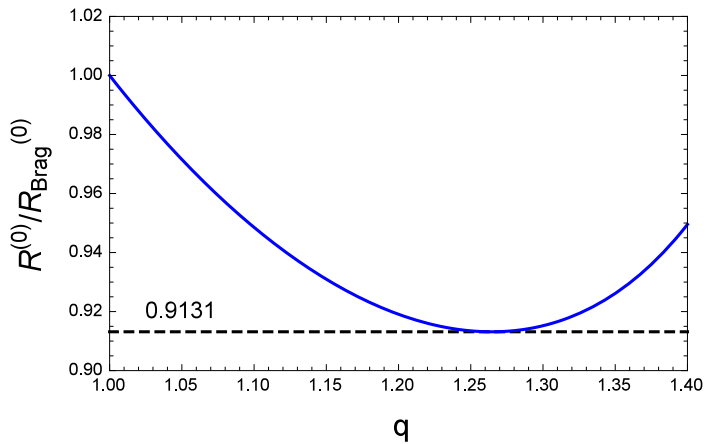}}
  \caption{The behaviour of $\mathbf{R}^{(0)}/\mathbf{R}_{Brag}^{(0)}$ as function of $q$. The decreases between $1<q\leq 1.26$ accounts the increasing number of suprathermal electrons which reduces  the cross section of the e-i collision. The growth after $q>1.26$ a consequence of the long-range correlations.}
\label{fig:friczero}
\end{figure}

The behaviour of  $\mathbf{R}^{(0)}/\mathbf{R}^{(0)}_{Brag}$, where $\mathbf{R}^{(0)}_{Brag}$, correspondents to $q\to1$ (where $A_q=1$ in this limit), as function of $q$ is depicted in figure \ref{fig:friczero}. The initial decrease is explained by the reduction of the e-i collision frequency $\omega_{ei}\propto v^{-3}$ (see Eq.(\ref{freqcolei})) due the increasing number of suprathermal electrons (faster electrons). After the minimum at $q\approx1.26$, the subsequent increase of $\mathbf{R}^{(0)}$ is understood as consequence of the long-range correlations, which are strong enough to overcome the reduction of $\omega_{ei}$ and increase the friction force, but without returning to the classical value. This behaviour suggests that, near $q=1.4$, the effect of the long-range correlations, although small, should be noticeable in transport phenomena. Then, the statistical description that includes the correlations between the fluid parameters are necessary for a rigorous analysis of the transport. However, this task is beyond of the scope of this work.

\section{Chapman-Enskog Method} \label{sec:CEmethod}

The solution of Eq.(\ref{eqkinpecv}) by the Chapman-Enskog (CE) method  is analogous to the classical procedure \cite{chapman,cercignani1990,Braginskii1965}. In the weak interaction limit, this solution is approximated by $f=f_0+f_1$; and $f_1/f_0 \ll1$. The direct substitution of $f$ in the referred equation leads to
\begin{eqnarray}
I_{ee}(f_0) + I_{ei}'(f_0) + \frac{e}{m} ( \mathbf{v}\times \mathbf{B}) \cdot \frac{ \partial f_0}{\partial \mathbf{v}} =0,
\label{eqkinzero}
\\
I_{e}(f_1) + \frac{e}{m} (\mathbf{v}\times\mathbf{B}) \cdot \frac{\partial f_1}{\partial t} = \frac{ d f_0}{d t} + \mathbf{v}\cdot \nabla f_0 + \left(\frac{e \mathbf{E}'}{m} + \frac{ d \mathbf{V}}{d t}\right) \cdot \frac{\partial f_0}{\partial \mathbf{v}} + C_{ei}'\left( \mathbf{v}\cdot \mathbf{U} f_0 \right),
\label{eqkinfirst}
\end{eqnarray}
where $I_e(f_1)=I_{ee}(f_1)+I_{ei}(f_1)$ are the linearised versions of the e-e and e-i collision operators and $C_{ei}'(\mathbf{v}\cdot \mathbf{U} f_0)$ is the small part of $C_{ei}'(f)$, all of them given in Appendix A. The separation of Eq.(\ref{eqkinzero}) and Eq.(\ref{eqkinfirst}) results from the ordination of the solutions imposed by the CE method.

The substitution of Eq.(\ref{distNtq}) into Eq.(\ref{eqkinzero}) proves that $f_0$ is the zero order solution. The zeroth order fluid equations, calculated from Eq.(\ref{eqkinpecv}) by taking the first three moments and supposing $f=f_0$, are
\begin{subequations}
\begin{align}
\frac{d n}{dt} + n \nabla \cdot \mathbf{V} =0;
\label{conteq0}
\\
n m \frac{ d\mathbf{V}}{d t} +\frac{ e \mathbf{E}'}{m} = \frac{ \mathbf{R}^{(0)} + \nabla p}{n m};
\label{moveq0}
\\
\frac{3}{2} \frac{ d T}{dt} + T \nabla \cdot \mathbf{V} =0.
\end{align}
\label{tempbaleq}
\end{subequations}

This set of equations, with exception of the explicit form of $\mathbf{R}^{(0)}$, Eq.(\ref{fric0}), is equal to the zeroth order fluid equations found in the classical model \cite{cercignani1990}. Similar fluid equations have been found from other transport models within q-statistics \cite{Boghosian1999,Potiguar2002}; however, none of them were introduced self-consistently as here. It is also important to notice that these equations correspond to the adiabatic evolution of the fluid (of course, subjected to the equation of state), which can be deduced only from thermodynamic and mechanical arguments. Since the Tsallis statistics does not change Newtonian mechanics \cite{Tsallis2009} and our temperature definition maintains its ordinary interpretation (see Eq.(\ref{gen0law})), the recovery of this system of equations is a test of self-consistency of our formulation.

The determination of the first order solution follows the standard procedure of \cite{chapman,Braginskii1965}; actually, it reduces to elimination of the time derivatives in Eq.(\ref{eqkinfirst}), with help of Eqs.(\ref{conteq0}-\ref{tempbaleq}), and reorganization of the remaining terms conveniently to find the solutions of the equation. In this reorganization, instead the usual associated Laguerre polynomials presented in the classical model, the right side of Eq.(\ref{eqkinfirst}) is written in terms of Jacobi polynomials \cite{jeffrey2008handbook}. This slightly change is almost self-evident and requires just a small amount of algebraic manipulation. 

The straightforward calculation of the first order kinetic equation yields
\begin{eqnarray}
I_{e}(f_1) + \frac{e}{m} (\mathbf{v}\times \mathbf{B}) \cdot \frac{\partial f_1}{\partial \mathbf{v}} &=& \left\{\left( 10 \frac{1-q}{1+q} - \frac{ 5 -3 q}{1+q} L_1^\frac{3}{2}\left(x^2;q\right)\right) \mathbf\cdot \nabla \ln T \right.\nonumber
\\\label{kineqlin1}
&& + \left[ 2 \frac{1-q}{1+q} \left( 10 \frac{1-q}{5 -3 q} + L_1^\frac{3}{2}\left(x^2;q\right)\right)\right] \mathbf{v}\cdot \nabla \ln p
\\\nonumber
&&\left. + \frac{ q \left( \mathbf{R}^{(0)}+\mathbf{R}^{(1)}\right)\cdot \mathbf{v}}{ m T_q}\right\}\frac{f_0}{1-(1-q)x^2}+ C_{ei}'\left(\mathbf{v}\cdot\mathbf{U} f_0\right)
\end{eqnarray}
where $L_1^{3/2}\left(x^2;q\right) = P_1^{(3/2,1/(1-q))}\left(x^2\right)= -5/2 + (1+q) x^2/2$ is the first degree Jacobi polynomial and $\mathbf{R}^{(0)}$ is given in Eq.(\ref{fric0}). In the above equation, the term proportional to $\nabla \ln p$ has no correspondent in the classical model (i.e., $q\to1$); it is an exclusive perturbation of q-statistics and originates from the non-cancellation between the terms provided by $\nabla f_0$ and $ \partial f_0/\partial \mathbf{v}$ due modification of the power-law distributions. This new transport term has already been identified in literature as an anomalous collisional transport \cite{Boghosian1999,Potiguar2002}.

The general solution of the linear equation, Eq.(\ref{kineqlin1}), can be written as a sum of the source terms on the right-hand-side, i.e.,
\begin{equation}
f_1 =  \left[A_T (x^2,q) \mathbf{v}\cdot \nabla \ln T+A_p (x^2,q) \mathbf{v}\cdot \nabla \ln p +A_U (x^2,q) \mathbf{v}\cdot \mathbf{U}\right] f_0,
\label{1solline}
\end{equation}
where the $A_j$'s are arbitrary functions.

The linear solution recovers the same bilinear relation between thermodynamic forces (perturbations) and associated (conductive) fluxes coupled by a transport coefficient, i.e., the well-know forms of the Fourier, Fick, and Ohm laws \cite{degroot1984}. In fact, this is straightforward verified from Eq.(\ref{defflux}) by the direct substitution of the general solution, where the transport coefficients are defined as integrals of the unknown $A$ functions. In particular, the transport coefficients of the heat flux due $\mathbf{U}$ and the friction force due $\nabla T$ are, respectively,
\begin{equation}
\alpha_U =\frac{ 5 -3q}{3 n(1+q)} \int d\mathbf{v} v^2 L_1^\frac{3}{2}\left( v^2;q\right) A_U f_0, \quad \alpha_T = - \frac{ q}{3 n T_q} \int d\mathbf{v} m v_\eta I_{ei}\left( v_\eta A_T\right), \label{thermeleccoef}
\end{equation}
where $I_{ei}$ is the linearised e-i collision operator given in Appendix A.

This allows the first order friction force to be defined, without loss of generality, by
\begin{equation}
\frac{q \mathbf{R}^{(1)}}{n T_q} = \alpha_w \mathbf{W}; \quad \mathbf{W} = (\nabla \ln T,\nabla \ln p, \mathbf{U}),
\label{genfric}
\end{equation}
where $\alpha_w$ is the corresponding transport coefficient to each perturbation ($w = T,p,U$). Hence, the linear relations in Eqs.(\ref{1solline}) and (\ref{genfric}) account for the separation of Eq.(\ref{kineqlin1}) in a distinct equation for each perturbation in $f_1$, as follows
\begin{eqnarray}
I_e(A_T \mathbf{v}) - i\Omega \mathbf{v} A_T f_0 = \left[ 10 \frac{1-q}{1+q} -\frac{ 5 -3 q}{1+q} L_1^\frac{3}{2}\left(x^2 ;q\right) + \alpha_T\right] \frac{ \mathbf{v} f_0}{1-(1-q) x^2},
\label{eqkinT}
\\
I_e(A_p \mathbf{v}) - i\Omega \mathbf{v} A_p f_0 = \left[ 2 \frac{1-q}{1+q}\left( 10 \frac{q-1}{5 -3 q} +L_1^\frac{3}{2}(x^2;q)\right)+\alpha_p\right] \frac{ \mathbf{v} f_0}{1-(1-q) x^2},
\label{kineqp}
\\
I_e(A_U \mathbf{v}) - i\Omega \mathbf{v} A_U f_0= \frac{ q(\eta_0 +\eta_1)}{nT_q} \frac{\mathbf{v} f_0}{1-(1-q)x^2} +C_{ei}'\left(\mathbf{v}f_0\right),
\label{kineqU}
\end{eqnarray}
where the perpendicular and diamagnetic equations are coupled by $A_w$ and $\alpha_w$ (the parallel direction ($||$) is obtained from the perpendicular taking $B\to0$), $A_w = A_w^\perp  + i\Omega A_w^\wedge$ and $\alpha_w =\alpha_w^\perp + i\Omega \alpha_w^\wedge$, $w=(T,p,U)$, $\Omega = e B/m$ is the cyclotron frequency, $\eta_0$ is the friction coefficient from Eq.(\ref{fric0}), and $\eta_1$ is the first order friction coefficient.

From the above set of equations and using the self-adjoint property of the collision operator from Eq.(\ref{app:selfad}) in appendix A, the following relations between the transport coefficients can be proved
\begin{equation}
\alpha_T = \alpha_U \equiv \alpha; \quad
\left(\begin{array}{ll}
\kappa_p \\
\alpha_p
\end{array}\right)
=
2\frac{q-1}{5 -3 q} 
\left(\begin{array}{ll}
\kappa_T \\\alpha_T
\end{array}\right)
\label{onsager}
\end{equation}
where $\alpha_T$ and $\alpha_U$ are given by Eq.(\ref{thermeleccoef}), and $\kappa_T$ and $\kappa_p$ are, respectively, the thermal conductivities due $\nabla T$ and $\nabla p$ calculated from Eq.(\ref{defflux}) as
\begin{equation}
\left\{
\begin{array}{c}
\kappa_T
\\
\kappa_p
\end{array}
\right\}
 = \frac{2}{3} \frac{T_q}{1+q} \int d  \mathbf{v} v^2 L_1^\frac{3}{2} \left(x^2;q\right) f_0
\left\{
\begin{array}{c}
A_T\\ A_p
\end{array}
\right\}.
\end{equation}

It is important to note that the transport coefficients of the convective fluxes in all magnetic directions are included in the above expressions due the coupling of the kinetic equations; they follow the same representation of $\alpha$ in Eqs.(\ref{eqkinT}), (\ref{kineqp}), and (\ref{kineqU}).

The identity between the coefficients in Eq.(\ref{onsager}) proves the Onsager reciprocity relations \cite{degroot1984}, as verified in other formulations of the q-statistics \cite{Chame1997,Caceres1995}, but without the explicit and self-consistent formulation presented here. In this context, the relations between $\kappa_T$ and $\kappa_p$, and $\alpha_T$ and $\alpha_p$ represent extended reciprocity relations, where all transport coefficients of the convective flux due $\nabla p$ are identified with those due $\nabla T$. Since the transport mechanism in our model is local (collisions), the new flux proportional to $\nabla p$ is understood as consequence of the local reorganization due the long-range correlations and driven in the same way as the ordinary heat flux induced by $\nabla T$. Furthermore, the extended reciprocity relations enable the coupling of the gradient driven forces leading to the transport matrix
\begin{equation}
\setlength{\arraycolsep}{0pt}
\renewcommand{\arraystretch}{1.3}	
\left(
\begin{array}{ll}  \displaystyle
\frac{\mathbf{q}_j}{nT}
\\  \displaystyle
\frac{q \mathbf{R}_j}{n T_q}
\end{array}\right)
= -
\left(
\begin{array}{cc}  \displaystyle
\chi_{jT} & \alpha_j\\  \displaystyle
\alpha_j &  \frac{ q (\eta_1 - \eta_0)}{n T_q}
\end{array}\right)
\left(
\begin{array}{lr}  \displaystyle
\nabla_j \ln \left(T p^{2 \frac{q-1}{5 -3 q}}\right)
\\  \displaystyle
-\mathbf{U}_j
\end{array}\right)\label{matrixtransp}
\end{equation}
where $\chi_{jT}$ is the heat diffusivity defined from $\kappa_{jT} = n \chi_{jT}$, $\mathbf{q}_j$ and $\mathbf{R}_j$ are the total heat flux and friction force, respectively, and the index $j$ stands for the parallel, perpendicular, and diamagnetic directions. In the above matrix notation, there is no diagonal term related to $\nabla p$; therefore, this driving force behaves as a non-diagonal term and, eventually, transport particles, energy, and momentum along or against $\nabla p$, as the thermoelectric fluxes for instance. Hence, the $A_p$ function could be defined up to a ``$\pm$'' sign, which results in the appearance of the same sign in the power law of $p$ in the generalized thermodynamic force. However, independently of this sign, the ordinary entropy production $\sigma_S \sim \mathbf{J}_a \cdot \mathbf{\mathcal{F}}$, where $\mathbf{J}_a$ is the convective flux and $\mathbf{\mathcal{F}}$ is the perturbation \cite{degroot1984}, is always positive, whether the direction of the flow is towards or against $\nabla p$; therefore, consistent with the second law of thermodynamics. This is a very important result from our model within the basic framework of irreversible transport, since in the previous models the direction of the irreversible fluxes violates this condition \cite{Boghosian1999,Potiguar2002}.

\section{Transport Coefficients} \label{sec:transcoe}

One of the most distinctive solution methods of the classical transport model for magnetized plasmas was introduce by Braginskii \cite{Braginskii1965}; basically, the first order solution of the kinetic equation is approximated by an asymptotic series of associated Laguerre polynomials and Maxwellian distributions. The main advantage of his method concerns the orthogonality relation of such polynomials, which simultaneously ensures the conditions of the CE-method and avoids the numerical solution of the first order kinetic equation. The same methodological principle can be adapted to our model with the appropriate modifications in the asymptotic expansion.

Following the method of Braginskii, the $A$ functions in the Eqs.(\ref{eqkinT}) and (\ref{kineqU}) are approximated by asymptotic series of Jacobi polynomials
\begin{eqnarray}
A_T = -\tau \sum_{k=1}^\infty \frac{a_k L_k^\frac{3}{2}\left(x^2;q\right)}{\left[1-(1-q)x^2\right]^{2 +k}},
\label{ATseries}
\\
A_U = \frac{m}{T_q} \sum_{k=1}^\infty \frac{a_k L_k^\frac{3}{2}\left(x^2;q\right)}{\left[1-(1-q)x^2\right]^{2+k}},
\label{AUseries}
\end{eqnarray}
where the coefficient $a_k = a_k^\perp + i\Omega a_k^\wedge$ are different for each series; the extended reciprocity relations obviate the necessity to solve the equation for $A_p$. 

The orthogonality properties of the power law asymptotic expansions are readily verified, for instance, when the orthogonal relation of the Jacobi polynomials \cite{jeffrey2008handbook} are employed in the conditions imposed by the CE-method
\begin{equation}
\int d\mathbf{v} \left( 1,m \mathbf{v}, \frac{m v^2}{2}\right) f_1 =0,
\label{CEcond}
\end{equation}
where $f_1$ is given in Eq.(\ref{1solline}) and the $A$ functions are, respectively, given by and Eqs.(\ref{ATseries}) and (\ref{AUseries}).

Taking advantage of the orthogonality of the Jacobi polynomials, the integral equations Eq.(\ref{eqkinT}) and Eq.(\ref{kineqU}) can be multiplied by the appropriated factor and integrated over $\mathbf{v}$, in order to obtain an infinity system of algebraic equations
\begin{equation}
\sum_{k=1}^\infty \left( c_{\ell k}^{ee} + c_{\ell k}^{ei} - i \Delta c_{\ell k}^B\right) = c_\ell, \label{algeq}
\end{equation}
where $\Delta = \Omega \tau$ and the matrix elements $c$ are given in appendix B; namely, $c_{\ell k}^{ee}$, $c_{\ell k}^{ei} $, $c_{\ell k}^B$, and $c_\ell$ correspond, respectively, to the integration of $I_{ee}$, $I_{ei}$, the magnetic term, and the source terms.

This equation system is similar to that found in the Braginskii model and, therefore, its solutions (the $a_k$ coefficients) exhibit the same asymptotic behaviour for strong magnetized plasmas $\Delta \gg1$. As it can be verified, the coefficients in the perpendicular and diamagnetic directions are proportional, respectively, to $\Delta^2$ and $\Delta$ (the parallel direction does not depend on $\Delta$).

Although this method allows the determination of all transport coefficients, here we will restrict the discussion only to the most relevant ones, namely, the perpendicular and parallel friction force coefficients, thermal conductivities, and the parallel thermoelectric coefficient. These coefficients are defined by the multiplicative factors of the thermodynamic forces at transport equation calculated from Eq.(\ref{defflux}); in terms of the asymptotic series given by Eqs.(\ref{ATseries}) and (\ref{AUseries}), such equations are
\begin{eqnarray}
&&\mathbf{R}=\mathbf{R}^{(0)}+\mathbf{R}_U^{(1)} \approx \mathbf{R}^{(0)} +\sum_{k=1}^\infty \frac{ n m}{\tau} r_U(q,k) a_k \mathbf{U}_{||}, \label{fricuasym}
\\
&&\mathbf{q}_U = \sum_{k=1}^\infty  n T\,t_U(q,k) a_k \mathbf{U}_{||}, \label{heatuasym}
\\
&&\mathbf{q}_T = -\sum_{k=1}^\infty \frac{ n T \tau}{n} c_T(q,k) \left[a_k \left(\nabla_{||} T + T \nabla_{||} \ln n\right)+ a_k'\left( \nabla_\perp T + T \nabla_\perp \ln n\right)\right],\qquad\label{heattasym}
\end{eqnarray}
where we have neglected the perpendicular component of $\mathbf{R}^{(1)}$ in the first expression, since it is proportional to $\Delta^{-2}$, whereas the same component in $\mathbf{R}^{(0)}$ is independent of $\Delta$, therefore, $\eta_\perp$ is given by Eq.(\ref{fric0}); we also have used $p= n T$ (see Eq.(\ref{defflux})) in the last equation. Then, $r_U$, $t_U$, and $c_T$ are defined by
\begin{eqnarray}
r_U(q,k) = \frac{ 2\pi^\frac{3}{2} A_q}{q^2} \left(\frac{2}{5 - 3 q}\right)^\frac{1}{2} \int_0^\infty dx\frac{x L_k^\frac{3}{2} ( x^2; q) \left[1-(1-q) x^2\right]^{\frac{q}{1-q}-1-k}}{\left[1-(1-q) x^2\right]^{1+k}},
\\
t_U(q,k) = \frac{2 \pi A_q}{1+q} \frac{ 5 - 3 q}{3} \int_0^\infty dx\frac{x^4 L_1^\frac{3}{2}(x^2,q) L_k^\frac{3}{2} (x^2,q) \left[1-(1-q) x^2\right)^{\frac{q}{1-q}}}{\left[1-(1-q) x^2\right]^{2 +k}},
\\
c_T(q,k) = \frac{ \pi(5 -3q)  A_q}{3 (1+q)} \int_0^\infty dx\, \frac{x^2 L_1^\frac{3}{2} (x^2;q) L_k^\frac{3}{2}(x^2;q) \left[1-(1-q)x^2\right]^{\frac{q}{1-q} }}{(3-q)^{-1}\left(1-(1-q)x^2\right)^{2 +k}}.
\end{eqnarray}

The coefficients $a_k$ of the asymptotic series are determined by Eq.(\ref{algeq}), where the truncation of the infinity system of equations represents the degree of the asymptotic series in Eqs.(\ref{ATseries}) and (\ref{AUseries}). All $c$ coefficients in these equations can be calculated analytically, except $c_{\ell k}^{ee}$, which is numerically obtained. This is a consequence of the power-law distribution not factoring in the integration variables as in the Braginskii model, for exponential-like distributions. This numerical calculation is performed for a list of predetermined values of $q$ using the Monte Carlo method with random and stratified sampling \cite{press1992numerical}.

In short, the predetermined list of nine values in the range $q\in[1.1,4]$, with $0.5$ pace, is evaluated in the analytical expressions of $c$ in Eq.(\ref{algeq}), whereas $c_{\ell k}^{ee}$ is calculated by the Monte Carlo method. Hence, the system of equations is solved by simple matrix inversion and the limit $\Delta\gg1$ is imposed, allowing the expansion of the $a_k$ as series of $\Delta$. As already mentioned, the dominant terms of this series are $\Delta^{-2}$ and $\Delta^{-1}$ for, respectively, the perpendicular and diamagnetic components; we also note that theses components are the real and imaginary part of $a_k = a_k^\perp + i \Omega a_k^\wedge$, as determined by the separation of the $A$ functions in Eqs.(\ref{ATseries}) and (\ref{AUseries}). The transport coefficients are then calculated from Eqs.(\ref{fricuasym}-\ref{heattasym}), where the factors $r_U$, $t_U$, and $c_T$ are also evaluated according to the predetermined list from its analytical expressions.

In order to easily understand the general behaviour of the transport coefficients, we also fit simple functions to the result and find the following expressions
\begin{eqnarray}
\eta_{||} &=&\left( 0.51 - \gamma_1 (q-1)^{0.8} \left[1-\gamma_2 ( 5-3q)^{1.5}\right]\right) \frac{ n m}{\tau}, \label{etapara}
\\
\alpha_{||} &=& 0.71 + \theta_1\frac{ (q-1)^{0.5}}{1+q} \left( 1- \theta_2 ( 3 -2 q)^{2}\right), \label{alphapara}
\\
\kappa_{||} &=&\frac{ n T\tau}{m} \frac{ 3 -q}{ 5 -3q} \left( 3.16 - 10 \xi_1 (q-1)^{1.5} \left[ 10 - \xi_2 \left(9 - 5 q\right)^{0.8}\right]\right), \label{kappapara}
\\
\kappa_{\perp} &=& \frac{ n T \tau}{m \Delta^2} \frac{ 3 -q}{5 -3q} \left( 4.66 + \xi_3 \frac{(q-1)^{1.25}}{(3-2q)^{0.5}} \left[ 10-\xi_4 (7 - 5q)^{1.1} \right]\right), \label{kappaperp}
\end{eqnarray}
where the fitted coefficients are given in table \ref{tab:fitcoefs}. In the limit $q\to1$, these expressions recover the transport coefficients of the Braginskii model, namely, 
\begin{eqnarray}
\eta_{||{Brag}}= 0.51 \frac{n m}{\tau}, &\qquad \qquad & \alpha_{||{Brag}}=0.71
\\
\kappa_{||{Brag}}= 3.16 \frac{n T \tau}{m} & \qquad \qquad & \kappa_{\perp{Brag}}=4.66 \frac{n T \tau}{m \Delta^2}
\end{eqnarray}

\begin{table}
  \begin{center}
  \begin{tabular}{|rc||c||rc||c||rc||c||rc|}
	\hline
   $\gamma_1$   &0.57  && $\gamma_2$ &0.13  && $\theta_1$   &4.85  && $\theta_2$ & 0.95 \\
   $\xi_1$      &1.46  &&	$\xi_2$    & 6.07 && $\xi_3$      &2.69 && $\xi_4$    &4.75\\
	\hline
  \end{tabular}
  \caption{Fitting values of the transport coefficients in Eqs.(\ref{etapara})}
  \label{tab:fitcoefs}
  \end{center}
\end{table}

\begin{figure}
  \centerline{\includegraphics{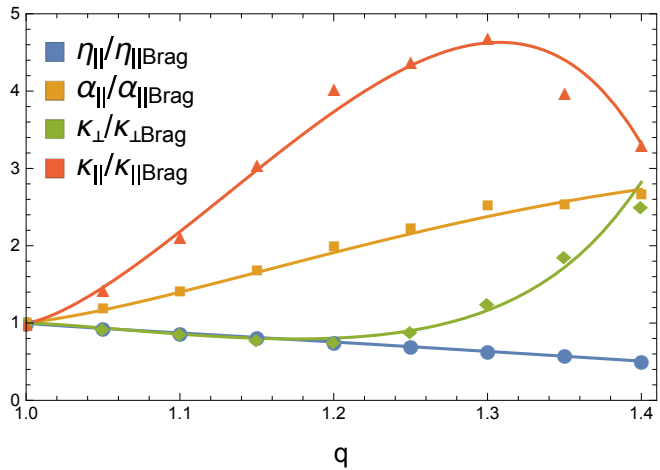}}
  \caption{Behaviour of the transport coefficient given in Eq.(\ref{etapara}) as function of $q$ and normalized by the Braginskii values. The markers are numerical evaluations of the the transport coefficients from Eqs.(\ref{fricuasym}-\ref{heattasym}) and the lines the fitted polynomials in Eqs.(\ref{etapara}).}
\label{fig:fitcoefs}
\end{figure}

The parallel friction force coefficient $\eta_{||}$ is calculated from Eq.(\ref{fricuasym}) and shown in figure \ref{fig:fitcoefs}, together with the other coefficients. In order to obtain the general behaviour of this coefficient, it is necessary to go up to the fifth order of the asymptotic approximation. Indeed, higher approximations are expected due the slow convergence of the power-law asymptotic series when compared with the classical model, where the expansion series are proportional to an exponential function. The $\eta_{||}$, therefore, $\mathbf{R}_{||}$, is a monotonically decreasing function of $q$, essentially because the e-e collisional relaxation is weakened by suprathermal electrons, enhancing the long-tail of the distribution. The expression Eq.(\ref{etapara}) gives a result with less than $\pm 5\%$ relative error up to $q=1.35$ and $\pm 8\%$ for $q=1.4$.

The decrease of the friction force with $q$ is particularly relevant for the mechanism of runaway electrons, since the threshold Dreicer field $E_D$, beyond which the electrons are accelerated indefinitely, are determined by the point of maximum friction force \cite{Helander2005}.

The parallel thermoelectric coefficient $\alpha_{||}$ is calculated from Eq.(\ref{heatuasym}) and the results of its numerical evaluation is presented in figure \ref{fig:fitcoefs}. Due to the slow convergence of the asymptotic series, we have to follow the approximation up to sixth order to ensure a reasonable convergence. Unfortunately, these calculations are very sensitive to numerical errors of the numerical integration of $c_{\ell k}^{ee}$. This problem is enhanced as the order of the approximation is increased, in special for values of $q\geq 1.2$. As a consequence, the relative error associated with this expression is smaller than $\pm 5 \%$ up to $q\leq 1.15$, but can vary from $\pm 7\%$ for $q=1.2$ to $25\%$ for $q=1.4$.

In spite of the mentioned calculation difficulties, the dependence of $\alpha_{||}$ with $q$ is quantitatively correct; as also pointed out in the appendix C. Indeed, its increase with $q$ shown in figure \ref{fig:fitcoefs} can be understood according the basic transport mechanism of the thermoelectric heat flux \cite{Helander2005,Hazeltine1998}: the net heat flow due the difference between faster and slower electrons moving, respectively, in directions $\mathbf{U}$ and $-\mathbf{U}$, are enhanced by the increasing number of suprathermal electrons (faster electrons) flowing along $\mathbf{U}$. The saturation/decreasing of the transport coefficient with $q$, as well as the thermoelectric heat flux, is due the long-range correlations.

The parallel thermal conductivity $\kappa_{||}$, evaluated up to sixth order of the asymptotic approximation, is given in Eq.(\ref{kappapara}). We warn that the accuracy of this expression deteriorates as $q\to1.4$, as in the case of $\alpha_{||}$; its is smaller than $\pm 8 \%$ for $q\leq1.15$ and can vary from $\pm 15\%$ for $q=1.2$ to $\pm 33\%$ for $q=1.4$. Again, such unsatisfactory variation for $q\geq 1.2$ is due to the high order of the Jacobi polynomials that are required as the value of $q$ increases, but the qualitative behaviour shown in figure \ref{fig:fitcoefs} is correct (see appendix C). In particular, the initial increase of $\kappa_{||}$ with $q$, due to the enhancement of the flux caused by the effect of suprathermal electrons, tends to saturate and eventually decrease as consequence of the long-range correlations.

Interestingly, the calculation of the heat transport coefficient across the magnetic field is much less sensitive to the error in the numerical calculation of $c_{\ell k}^{ee}$. In this case, the curves resulting from the different orders of calculations alternate with respect to an average one, so that the one corresponding to the sixth order is reasonably precise up to $q\approx 1.25$. The expression for $\kappa_\perp$ up to the sixth order of approximation is given by Eq.(\ref{kappaperp}) and it is represented in figure \ref{fig:fitcoefs}. The relative error associated with this expression is smaller than $\pm 5\%$ for $q\leq 1.2$ and can vary from $\pm 6\%$ for $q=1.25$ to $9\%$ for $q=1.4$. It is evident from the figure that $\kappa_\perp$ initially decreases as the tail of the electron distribution function enlarges, up to $q\approx1.2$. Above this value, the heat transport coefficient increases again, even beyond the value for the Braginskii model, corresponding to $q\to 1$.

\begin{figure}
  \centerline{\includegraphics{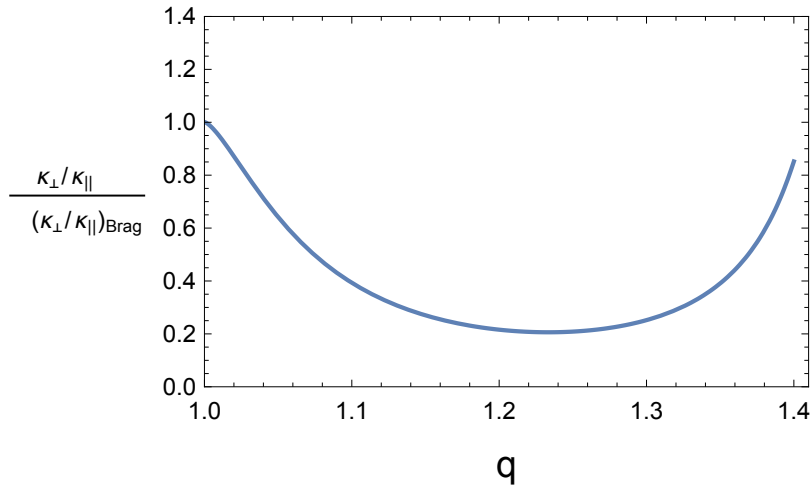}}
  \caption{ $\kappa_\perp/\kappa_{||}$ as a function of $q$ and normalized by the correspondent frequency of the classical model. The non-monotonicity is due the competition between the enhancement of the transport by suprathermal particles and the suppression of the local transport mechanism due the long-range correlations.}
\label{fig:ratiofreq}
\end{figure}

Independently of the statistical distribution, the general behaviour of the ratio between the perpendicular and parallel heat transport can be inferred from basic mechanical arguments, as $\kappa_\perp/\kappa_{||} \sim \omega^2/\Omega^2$, where $\omega$ is sort of a characteristic parallel collision frequency, i.e., the frequency of the scattering process presented by the collision operator without the dynamic effects of the evolution of $f$ and the magnetic field. For the standard Braginskii model, it can be shown that $\omega \sim \tau^{-1}$, where $\tau$ is given by Eq.(\ref{taurelax}) \cite{Helander2005}. In the same sense, the normalized ratio $\kappa_\perp/\kappa_{||}$ obtained from Eqs.(\ref{kappapara}) and (\ref{kappaperp}) is plotted in figure \ref{fig:ratiofreq}.  The initial decrease can be attributed to the weakening of the scattering process due to the suprathermal electrons. Then, after reaching a minimum, the ratio starts to increase due to the effect of the long-range correlations, which shows the suppression of the short-range correlation scattering process (collisions) in counterpart to the effect of long-range correlations.

Lastly, it is important to note that the increase of $\kappa_\perp$ above the classical value as $q\to1.4$ is a direct consequence of the suprathermal electrons, responsible for the new component of the heat flux carried by $\nabla p$, and represented by the multiplicative factor $(3 -q)/(5 -3 q)$ in Eqs.(\ref{kappapara}) and (\ref{kappapara}).

\section{Applications}

\subsection{Heat Flux in the Solar Wind}

In the solar wind, the measurements of the field-aligned electron heat flux are not fully consistent with the predictions from the classical transport models, due the presence of the suprathermal particles \cite{Salem2003}. Nevertheless, empirical models generalizing the classical heat transport by addition of a convective term accounting for the suprathermal electrons have been reproduced the data successfully \cite{Bale2013}. For instance, one of the most successful is the Hollweg model \cite{Smith2012}
\begin{equation}
\mathbf{q}_{||} = - \kappa_{Brag} \nabla T + \frac{3}{2} n T \alpha_H \mathbf{V},
\end{equation}
where $\alpha_H$ is the Hollweg constant and $\mathbf{V}$ the solar wind speed. This model generalizes the classical heat transport equation by assuming that part of the total heat flow is carried by a convective flux, when the flow velocity is comparable to the sound speed or when the electric potential that permeates the plasma is of order of the Dreicer field \cite{Landi2003}. In the sequel, we will show that this convective term is an effect of the suprathermal electrons.

From Eqs.(\ref{matrixtransp}), (\ref{alphapara}), and (\ref{kappapara}), we find
\begin{equation}
\mathbf{q}  = - \kappa \nabla T + 2 \frac{q-1}{3-q} \kappa e \mathbf{E} + \alpha n T \mathbf{U}, \label{hollqmodel}
\end{equation}
where $\nabla \ln n = - e \mathbf{E}/T$ from Eq.(\ref{nboltzmann}) and the index $||$ was suppressed. If the electric field approaches the Dreicer field $E_D \sim m v_T/(\tau e)$, which is consistent with $U=V\sim v_T$ \cite{Marsch2006,Jiulin2013}, the above equation yields
\begin{equation}
\mathbf{q}=-\kappa \nabla T + \left[ 2 \frac{q-1}{q-3} \frac{ m \kappa}{n T \tau} + \alpha\right] n T \mathbf{V}, \label{hollapprox}
\end{equation}
where the square brackets defines the Hollweg constant
\begin{equation}
\alpha_H = 2 \frac{q-1}{q-3} \frac{m \kappa}{n T\tau} +\alpha.
\end{equation}

\begin{figure}
  \centerline{\includegraphics{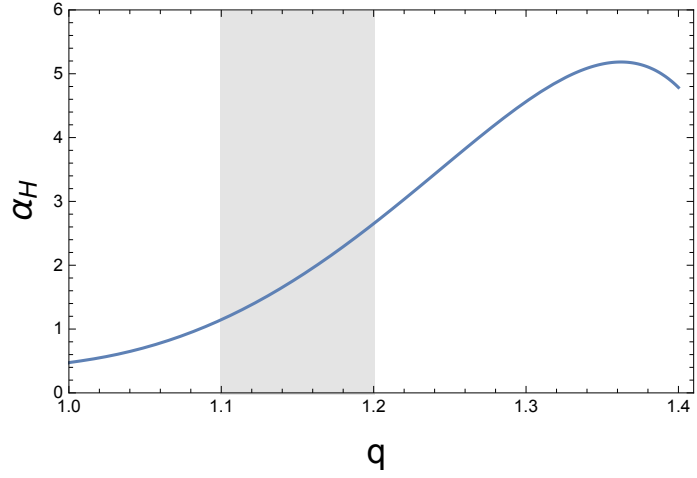}}
  \caption{Hollweg constant as function of $q$ including thermoelectric transport.}
\label{fig:hollwegcte}
\end{figure}

The observed $\kappa$-distributions correspond to $q=1.1-1.5$ and $\alpha_H =0.5 -10$ \cite{Bale2013,Pierrard2010,Landi2012}, both consistent with the result of $\alpha_H$ depicted in figure \ref{fig:hollwegcte}. In particular, typical expected values of the constant are $\alpha_H = 0.5-2$, for $q=1.1-1.2$ \cite{Bale2013,Smith2012,Landi2003}, which are quite close to our predicted result indicated by the shaded area in the figure. Even more accurate results are expected from the direct numerical calculation using Eq.(\ref{hollqmodel}), instead of the approximated Eq.(\ref{hollapprox}).

\begin{figure}
  \centerline{\includegraphics{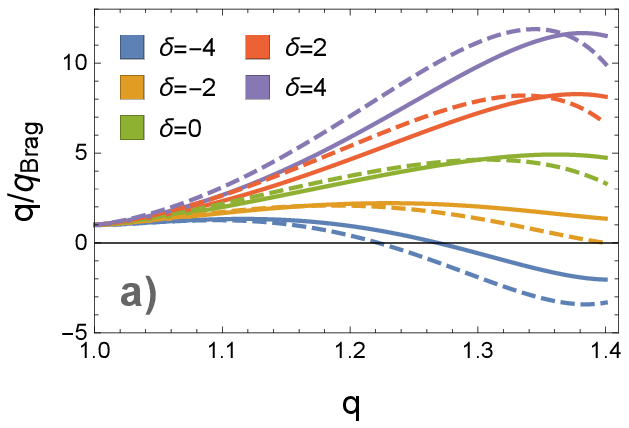}\hspace{.3cm}
							\includegraphics{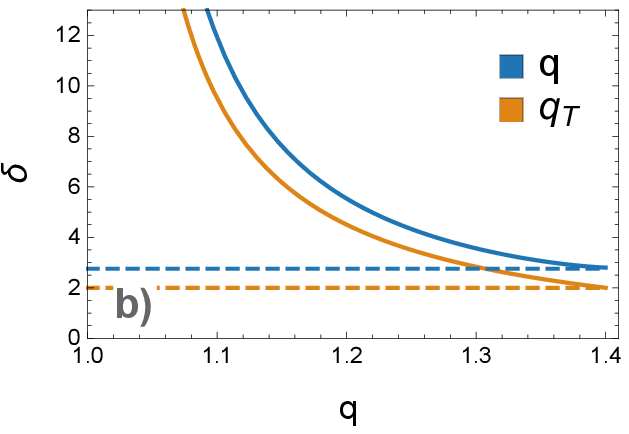}}
  \caption{Behaviour of the of the heat flux $\mathbf{q}$ as function of $q$. a) The inversion of the direction of the flux due the velocity filtration effect is noticeable. The dashed lines is the heat flux without thermoelectric effects. b) The minimum value of $\delta$ which vanishes the heat flux given by Eq.(\ref{hollqmodel}). The $q_T$ is the heat flux without the thermoelectric flux.}
\label{fig:velfil}
\end{figure}

Equation \ref{hollqmodel} also displays the so-called velocity filtration effect; a trapping effect on thermal electrons due the induced polarization of the local plasma potential by suprathermal particles capable of reverse the direction of the heat fluxparticles \cite{Scudder1992,Dorelli1999}. Figure \ref{fig:velfil} shows the heat flux for different $\delta =  | e \mathbf{E}|/|\nabla T|$, where Eqs.(\ref{moveq}), (\ref{matrixtransp}), (\ref{etapara}), and (\ref{kappapara}) were used, supposing a stationary plasma. The trapping effect is shown when $\mathbf{E}$ is anti-aligned to $\nabla T$, and the total heat is reduced or even reversed. From the figure \ref{fig:velfil}.b, we see that the inversion of the heat flux requires a polarization effect such that $\delta \approx e \phi/T >1$; even when the thermoelectric heat flux is neglected as shown by the dashed line. Therefore, it is not expected to occur within the conductive transport, where it is assumed that $e \phi/T \ll1$.

\subsection{Cold Pulse in Tokamaks}

A striking phenomenon observed in tokamak plasmas (and sterellators) is the temperature increase in the plasma core as response to a temperature reduction at the edge due pellet injection (impurities) \cite{Gentle1995,Callen1997}. This phenomenon is usually explained as an inward heat flux (heat flowing toward $\nabla  T$), what, in principle, contradicts the second law of thermodynamics, which states that the irreversible flux has to be against $\nabla T$ to maximize the entropy \cite{degroot1984}.

The phenomenon is commonly explained by empirical diffusive models of heat transport divided in two major groups: those that assume a critical temperature gradient that allows for an abrupt change of the thermal transport coefficient \cite{Kinsey1998,Neu2000}; and those that assume an additional convective term in the expression for the total heat flux \cite{Gao2014,Zou2000}. These modifications are understood as consequence of long-range transport mechanisms, since the characteristic time estimated for the transport phenomena in the experiments is smaller than the diffusion time \cite{Gao2014,Ryter2000}. Recently, the local transport paradigm was reintroduced from simulation results, which reproduced qualitatively the transport phenomenon. However, the edge perturbation, the characteristic time, the amplitudes for the flux and temperature were off when compared to the data (not even considering other unrealized predictions) \cite{Rodriguez2018}.

From Eqs.(\ref{moveq}) and (\ref{fric0}), neglecting thermoelectric effects due to the strongly magnetized plasma condition, $\nabla_\perp p \approx -e \mathbf{E}$, we find for a stationary plasma that $\mathbf{U}_\perp$ is given by
\begin{equation}
\mathbf{U}_\perp = - \frac{e^2 \eta_{0\perp}}{m^2 \Omega^2} \nabla_\perp p.
\end{equation}

Substituting the above result in the heat flux at Eq.(\ref{matrixtransp}) yields
\begin{equation}
\mathbf{q}_\perp = - \frac{5 -3q}{3 -q} \kappa \nabla_\perp T + 2 \frac{q-1}{3 -q} \frac{m^2 \Omega^2 \kappa_\perp}{n e^2 T \eta_0} \mathbf{U}_\perp,\label{coldpulse}
\end{equation}
recovering the exact form of the non-local transport empirical models \cite{Gao2014}. We also note that the ratio $\kappa_\perp/\eta_0$ is independent of $\tau$, which can be wrongly interpreted as a convective flux. Therefore, according to our formalism, the source of the inward heat flux is the presence of an extended tail in the distribution function, i.e., $q>1$, due suprathermal electrons.

We also point out that critical gradient models can, in principle, be recovered from our formalism if $\nabla p$ is included in the definition of $\kappa$ in the heat flux in Eq.(\ref{matrixtransp}). If we further suppose that the parameter $q\equiv q(\mathbf{r})$ could abruptly change, for instance due to some instability that modifies the electron distribution, the behaviour predicted by the critical gradient models for the heat conductivity could be approximated. This hypotheses is supported by the off-axis electron cyclotron resonant heating (ECRH) experiments, where the observed rise of the temperature in the center of the plasma is related to the same transport mechanism as in the cold pulse phenomena \cite{Neu2000,Ryter2000}. Moreover, the similarity between the mechanism underlying such effects is further depicted in the investigations of the electron heat transport in Tore Supra, where the general form of the convective models (see Eq.(\ref{coldpulse})) were able to reproduce the data\cite{Song2012}. 

In the off-axis ECRH experiments, the current accepted explanation for the temperature increasing in the plasma core is the abrupt change of the diffusivity $\chi$, when the radial profile of $\nabla T/T$ exceeds the trapped electron mode (TEM) threshold \cite{Ryter2003,Ryter2005}. In our model, such behaviour should be consequence of a sudden production of suprathermal electrons. In fact, the presence of these particles were observed associated with the off-axis ERCH \cite{Ryter2003}; in particular, instabilities due trapped or barely trapped electrons modes caused by suprathermal electrons were also identified \cite{Wong2000,Chen2009}.

Although our simple model prevents direct comparison with tokamak data, since neoclassical and potential fluctuations (turbulence) effects are not included, the entire cold pulse dynamic is understood just from Eq.(\ref{coldpulse}). The cold pulse phenomenon begins with impurity injection followed by the fast electron cooling at the plasma edge. This process is know to produce suprathermal electrons due the low collisionality of the particles from the initial non-perturbed distribution tail, which cannot thermalize into the new local equilibrium state \cite{Helander2004}. The self-consistent heat transport is then given by Eq.(\ref{matrixtransp}), where the second term is an exclusive contribution of the suprathermal electrons that cannot be achieved within local transport models supposing Maxwellian distributions. Hence, as mentioned before, this new component behaves like a non-diagonal term and could transport heat in favor of $\nabla T$, i.e., the heat flux inward. The suprathermal electron population necessary by our model at the plasma edge was identified in discharges in Tore Supra with impurity injection \cite{Zou2000}.

It is worth to mention that the suprathermal electrons could also enhance the effect of the noninductive current in tokamaks \cite{kadomtsev2012reviews}. In this effect, an abrupt increase of the current density in the radial (perpendicular) direction at the plasma edge produces an inductive electric field in opposition to the perturbation in order to maintain the total flux constant. This is then followed by diffusive spreading of the induced current reducing the profile of the plasma current density, except in the region of the perturbation, which remains above the unperturbed value. In other words, the resistivity is reduced by the polarization effect in the whole plasma column, except in the region of the perturbation. Hence, as consequence of the reduction of the perpendicular transport, when the total current is maintained constant, the temperature in the plasma core rises. As seen in figures \ref{fig:friczero} and \ref{fig:fitcoefs}, around $q\sim 1.15$, both $\kappa_\perp$ and $\eta_\perp$ (see Eq.(\ref{fric0})) are reduced, therefore, the inward heat flux is increased whereas the outward effective flux is reduced. This, of course, enhances the aboce process and could explain the core temperature rise in the context of the first analysis of the off-axis ECRH \cite{Petty1994}. However, it is difficult to numerically specity the contribution of the noninductive current in the rise of the core temperature in a general discussion.

\section{Discussion and Conclusions}

In this work, we present a kinetic model based on non-extensive statistics capable of determining the fluid transport equations. Starting just from the definition of the $S_q$, we were able to derive the equilibrium distribution function, the equilibrium temperature through the generalized zeroth law of thermodynamics, and the kinetic equation with the q-Landau operator, the consistent collisional operator for the weak interaction between charged particles. The derivation was kept as general as possible, ensuring all necessary conditions for a feasible kinetic model. This is also guaranteed by that, despite our further restriction on the range $1<q<5/3$, the model holds for the whole range of $q$ ($-\infty<q<5/3$).

As an practical application, we derived the fluid equations for the electrons in strongly magnetized plasmas. These calculations were carried out by the Chapman-Enskog method, where the solutions are approximated up to the first order, $f\approx f_0+f_1$. For the zero order solution, the $\mathbf{R}^{(0)}$ is calculated in Eq.(\ref{fric0}) and the result is depicted in figure \ref{fig:friczero}. The non-monotonic behaviour is understood as the competition between two effects: the decrease of the friction force as consequence of $\omega_\mathrm{ei}\sim v^{-3}$ due the increase number of suprathermal electrons; and, after the minimum at $q\approx 1.26$, the increasing due the long-range correlations, which overcome the suprathermal effect.

Using only the general aspects of the $f_1$ solution, provided by the Chapman-Enskog method, we also proved the Onsager reciprocity relations as weel as introduced the extended reciprocity relations (see Eq.(\ref{onsager})). These new relations identify the transport coefficients associated with the fluxes of $\nabla p$, a driven force exclusive of the q-kinetic theory, with those of $\nabla T$. This allows us to rewrite the transport equations in a matrix form in Eq.(\ref{matrixtransp}) by generalizing the gradient thermodynamic force as $\nabla \ln ( T p^{2(q-1)/(5 -3q)})$. In particular, this formulation guarantees the positiveness of the ordinary entropy production, even if the flow is in favor of $\nabla p$. This was a recurrent problem faced by the previously q-kinetic models \cite{Boghosian1999,Potiguar2002}, since they enable the existence of negative transport coefficients, which correspond to a sink of entropy in contradiction with the second law of the thermodynamics.

The $f_1$ in Eq.(\ref{1solline}) was approximated in Eqs(\ref{ATseries}) and (\ref{AUseries}) by asymptotic series of Jacobian polynomials. Such particular choice was made to take advantage of the orthogonal properties of these especial polynomials and ensures the Chapman-Enskog conditions. This also enables the transformation of the first-order kinetic equation into a system of algebraic equations, which is used to determine the coefficients $a_k$ of the asymptotic expansion and, therefore, the calculation of the transport coefficients. Due the characteristic power-law distributions, the asymptotic expansion have to be carried out until the fifth or the sixth order to guarantee reasonable accuracy for the transport coefficients. Except for the $\mathbf{R}_{||}$, all other calculated transport coefficients show a non-monotonic behaviour due the competition of the suprathermal and long-range correlation effects (see figure \ref{fig:fitcoefs}). This behaviour is readily understand by $\kappa_\perp/\kappa_{||} \sim \omega^2/\Omega^2$, normalized by the Braginskii transport coefficients, depicted in figure \ref{fig:ratiofreq}. The scattering process presented by the collision operator are enhanced by the suprathermal electrons until $q\approx 1.2$, where it starts to reduce due the long-range correlations. This behaviour is in accordance with the usual hypotheses of the weak turbulence models, which neglect in general the short-range transport mechanism (collisions) due its unimportance in the transport when long-range correlations are strong \cite{balescu2005}.

The presented transport equations were also applied in heat transport for the solar wind and in the cold pulse in laboratory plasmas. Our model was able to recover the empirical Hollweg model for the heat transport in solar winds \cite{Smith2012}. We rigorously identified that the part attributed to the convective flow originates from the conductive flux associated with the suprathermal electrons. The numerical values of the Hollweg constant ($\alpha_H$) shown in figure \ref{fig:hollwegcte} are consistent with the results found in the literature \cite{Bale2013,Smith2012,Landi2003}. In figure \ref{fig:velfil}, we also show that inversion of the heat flow direction by the velocity filtration effect requires an electric field much stronger than the one supported by the weak interaction assumption. The whole dynamic of the cold pulse could be understood only through Eq.(\ref{coldpulse}), which recovers the exact form of the empirical convective models. In our model, the inward heat flux is consequence of the suprathermal electrons, produced by the fast cooling of the plasma edge due impurity injection. This interpretation is supported by experimental evidence of suprathermal electrons in the plasma edge after the impurity injection \cite{Zou2000}.

In summary, in this work, for the first time, as far as we know, a self-consistent transport model for the non-extensive kinetic theory was presented. The general methodology was rigorously developed and the transport equations consistent with suprathermal electrons in a strong magnetized plasma were obtained. This simple model was applied in two poorly understood plasma phenomena, showing the importance of the suprathermal electrons in space and laboratory plasmas. We hope that the theoretical findings presented here could help to improve the actually understanding and description of the suprathermal electrons.

\section*{Acknowledgement}

This work is supported by the National Council of Scientific and Technological Developments (CNPq), under grants 307984/2016-8 and 157898/2014-8. We would like to thank professors Dr. C. Tsallis, Dr. A. Elfimov, and Dr. G. P. Canal for useful discussions.

\appendix
\section{Properties of the collision operator}\label{appA}

In the weak interaction assumption, we have $f_1/f_0\ll1$, therefore, if $f_1 = f_0 \psi$ then $\psi\ll1$. Applying this condition for electron-electron collision operator Eq.(\ref{qlandopab}) in the weak interaction assumption, the linearised version of the collision operator is
\begin{eqnarray}
I_{ee}&=& C_{ee}(f_0,f_0'\psi') + C_{ee}(f_0 \psi,f_0')\nonumber = \frac{2\pi e^4\lambda}{m} \frac{5 -3 q}{2q^2} \frac{ \partial}{\partial v_{\nu}} \times
\\
 &&\int d\mathbf{v}'f_{0}f_{0}' U_{\mu\nu} \left[ \frac{\partial}{\partial v_{\mu}} \left(\left[1-(1-q)x^2\right]\psi\right)- \frac{\partial}{\partial v_\mu'}\left[\left(1-(1-q)\psi'\right)\right]\right].\label{app:linopee}
\end{eqnarray}

The linearised version of the electron-ion collision operator is defined as the principal part of the collision operator (see Eqs.(\ref{opei}) and (\ref{prinopei})).
\begin{equation}
I_{ei} = \frac{2 \pi e^2 e_i^2\lambda n}{m^2} \frac{\partial}{\partial v_\nu}\left[ V_{\mu\nu} \frac{5-3q}{2 q^2} f_0 \frac{\partial}{\partial v_\mu} \left[(1-(1-q)x^2\right] \psi\right] \label{app:linopei}
\end{equation}
where $ x^2 =mv^2/2 T_q$ and $V_{\mu\nu} \equiv U_{\mu\nu}(\mathbf{v}_i=0)$.

If we define
\begin{equation}
\hat{f}=\left[1-(1-q)x^2\right] \psi, \qquad \hat{f}'=\left[1-(1-q){x'}^2\right] \psi',
\end{equation}
the self-adjoint property of the collision operator \cite{Helander2005} in Eq.(\ref{app:linopee}) can be proved when this equation is multiplied by $\hat{g}$ and then integrated over $\mathbf{v}$. Since in this circumstance both integration variables are dubbed, we can change $\mathbf{v}\to \mathbf{v}'$ and recover the same result. Hence, the self-adjoint property of the collision operator is expressed as
\begin{equation}
S_{ee}[\hat{f},\hat{g}]=\frac{2 \pi e^4 \lambda}{m^2} \frac{ 5 -3q}{2 q^2} \int d\mathbf{v} d\mathbf{v}' f_0 f_0' U_{\mu\nu} \left(\frac{\partial \hat{g}}{\partial v_\nu} -  \frac{\partial \hat{g}'}{\partial v_\nu'}\right) \left( \frac{\partial \hat{f}}{\partial v_\mu} - \frac{\partial \hat{f}'}{\partial v_\mu'}\right)=S_{ee}[\hat{g},\hat{h}], \label{app:selfad}
\end{equation}
where we can see the symmetry between the exchange of functions $\hat{f}$ and $\hat{g}$.

The proof of such symmetric relation for Eq.(\ref{app:linopei}) is trivial, since the operator is linear in $\hat{f}$, therefore,
\begin{equation}
S_{ee}[\hat{f},\hat{g}]=-\frac{2 \pi e^4 \lambda}{m^2} \frac{ 5 -3q}{2 q^2} \int d\mathbf{v} d\mathbf{v}' f_0 f_0' V_{\mu\nu}  \frac{\partial g}{\partial v_\nu} \frac{\partial \hat{f}}{\partial v_\mu}.
\end{equation}

In order to separate the small part of $C_{ei}'(f)$ in Eq.(\ref{eqkinfirst}), we can add and subtract at the full e-i collision operator in Eq.(\ref{qlandopab}) an ion distribution function shifted such that the mean ion velocity coincides with the mean electron velocity, just as in the Braginskii model \cite{Braginskii1965},
\begin{equation}
C_{ei}(f,f_i) = C_{ei}'(f,f_i')+C_{ei}'(f,f_i-f_i').
\end{equation}

The first term on the right-hand-side of the above equation is independent of $\mathbf{V}_i$; therefore, it is approximated by Eq.(\ref{prinopei}); the other term is the small term that can be by approximated by the zeroth order solutions, i.e., $f_0$. Since this difference is small in this order, it can be expanded in power series of $\mathbf{U}$, which recovers the expression of e-i collision operator for $\mathbf{R}^{(0)}$ with the opposite sign, that is,
\begin{equation}
C_{ei}'(f,f_i-f_i') = - \frac{2 \pi e^2 e_i^2 n \lambda}{m^2} \int d\mathbf{v} \frac{\partial}{\partial v_\nu} \left[U_{\mu\nu}\frac{\partial f_0^*(\mathbf{v} -\mathbf{U})}{\partial v_\mu}\right] = -C_{ei}'( \mathbf{v}\cdot \mathbf{U}f_0),
\end{equation}
which is also independent of $\mathbf{V}_i$. This expression is the same as in the Braginskii model, except by $f_0^*$.

\section{Coefficients of the algebraic equation} \label{app:coefalg}

The coefficients $c$ of the integral transformation of the kinetic equations Eq.(\ref{eqkinT}) and Eq.(\ref{kineqU}) are equal, except by the term on the right side. Their expressions are
\begin{eqnarray}
c_{\ell;U} = - \int d\mathbf{v}\left( \frac{4}{15} \frac{\tau}{2n} \frac{v_\xi}{\left[1-(1-q)x^3 \right]^{\ell +1}}\right) L_\ell^\frac{3}{2}\left(x^2;q\right) C_{ei}'\left( v_\xi f_0\right); \label{app:RclU}
\\
c_{\ell; T}= \frac{5 -3q}{1+q} \frac{16 \pi}{15} A_q \int_0^\infty dx x^4 L_1^\frac{3}{2}(x^2;q) L_k^\frac{3}{2}(x^2;q)[1-(1-q)x^2]^{\frac{q}{1-q}-2-\ell};
\\
c_{\ell k}^{m} = \frac{ 8\pi}{15} A_q \int d\mathbf{x} \frac{ x^2 L_\ell(x^2;q) L_k^\frac{3}{2}(x^2;q)}{[1-(1-q)x^2]^{3+\ell+k}} [1-(1-q)x^2]^\frac{q}{1-q};
\\
c_{\ell k}^{ei} = \frac{2 A_q \pi^\frac{3}{2}}{5q^2} \left(\frac{2}{5-3q}\right)^\frac{1}{2}  \int_0^\infty dx \frac{x L_\ell^\frac{3}{2} (x^2;q) L_k^\frac{3}{2}(x^2;q)[1-(1-q)x^2]^{\frac{q}{1-q}}}{[1-(1-q)x^2]^{2 +k+ \ell}};
\\
c_{\ell k}^{ee} =  \int d\mathbf{v} \left( - \frac{4}{15} \frac{1}{n} \frac{m}{2 T_q} \frac{\mathbf{v} L_k^\frac{3}{2}(x^2;q)}{[1-(1-q)x^2]^{1+\ell}}\right) I_{ee}(f_0,f_1),\label{app:cee}
\end{eqnarray}
where the terms inside the parentheses in Eq.(\ref{app:RclU}) and Eq.(\ref{app:cee}) are the multiplicative factor used in the transformation of the kinetic equation, the collision operators used in $c_{\ell k}^{ee}$ and in both $c_{\ell k}^{ei}$ and $c_{\ell;U}$ are given, respectively, by Eq.(\ref{app:linopee}) and Eq.(\ref{app:linopei}).

The integrals for all coefficients, but $c_{\ell k}^{ee}$, can be analytically evaluated. The integral of $c_{\ell k}^{ee}$ in numerically evaluated by the Monte Calor method from Eq.(\ref{app:cee}).

\section{Monte Carlo method and numerical error}

The choice of the Monte Carlo method is due to the high dimension of the integral, which is not well approached by quadrature techniques \cite{press1992numerical}. The optimization routines of the method were also chosen in accordance with the computation performance. The random sampling with stratification of each axis in 4 subdivisions shown shorter time and smaller error in comparison with other routines. In particular, the adaptive techniques as well as the importance sampling were inefficient due symmetries of the integrated function and the absence of regions of high accumulation (in the sampling phase of the method) when $q\to1.4$.

The stratified Monte Carlo method divides the axis of the six integration variables in four parts, totalling 1296 subspaces, and samples approximately 8 millions of points in each of the 30 rounds of integration, for each of 9 values ranging over $q\in[1.1,4]$ with pace of $0.5$. The numerical error is then estimated by the standard deviation of this collection. As the error provided by the method, the standard deviation is understood as a probability range where the absolute numerical error could be found \cite{press1992numerical}. Since the convergence of the Monte Carlo method is $\sim 1/\sqrt{N}$, where $N$ is the number of points sampled, if the $N$ was quite large, the estimated value of the integral is well estimated by the mean, even if the relative error of the integral is inaccurate.

The sensibility of the transport coefficients with the numerical error of the integration is partly due to the  the high order polynomials resulting from the solution of the algebraic equations in Eq.(\ref{algeq}). The other part is directly related to the form of the transport coefficient; for instance, $\alpha_{||}$, and $\kappa_{||}$, respectively, Eqs.(\ref{alphapara}) and (\ref{kappapara}), uses the same $a_k$, besides their error are different. Therefore, since only the error source is the numerical evaluation of $c_{\ell k}^{ee}$, this difference is addressed to through the propagation of the error in their definitions (see Eqs.(\ref{heatuasym}) and (\ref{heattasym}). 

We also note that even a small imprecision could cause large differences due the mixing of $c_{\ell k}^{ee}$ with very different scales. For $q=1.35$, the integration via Monte Carlo method results in $c_{1 6}^{ee} \approx 0.000039$ and $ c_{66}^{ee} \approx 1.19101$. Therefore, an insignificant variation in $c_{66}^{ee}$ could be enough to overcome the importance of the $c_{16}^{ee}$. In fact, this is was verified in the calculations of the sixth order approximations for the transport coefficients in Eqs.(\ref{alphapara}), where the increase of the precision in the lower order matrix element, for example, $c_{34}^{ee}$, was more effective in reducing the overall error when the precision is increased in the high order elements as $c_{1 6}^{ee}$.

\bibliographystyle{elsart-num}

\bibliography{main}

\end{document}